\documentclass[10pt]{article}

\usepackage{amssymb,amsthm,amscd, amsbsy, array}

\let\ssection=\section
\renewcommand{\section}{\setcounter{equation}{0}\ssection}

\begin{document}


\title{
Celestial Mechanics, Conformal Structures,\\
and
Gravitational Waves
}

\author{
C. DUVAL\footnote{mailto: duval@cpt.univ-mrs.fr}\\
Centre de Physique Th\'eorique, CNRS, 
Luminy, Case 907\\ 
F-13288 Marseille Cedex 9 (France)\footnote{ 
UMR 6207 du CNRS associ\'ee aux 
Universit\'es d'Aix-Marseille I et II et Universit\'e du Sud Toulon-Var; 
Laboratoire affili\'e \`a la FRUMAM-FR2291
}
\and
Gary GIBBONS\footnote{mailto: G.W.Gibbons@damtp.cam.ac.uk}\\
Centre for Mathematical Sciences,
Wilberforce Road,\\
Cambridge CB3 0WA
(United Kingdom)
\and
P.~A.~HORV\'ATHY\footnote{mailto: horvathy@lmpt.univ-tours.fr}\\
Laboratoire de Math\'ematiques et de Physique Th\'eorique\\
Universit\'e de Tours, Parc de Grandmont\\
F-37200 Tours (France)
}

\date{Phys. Rev. D {\bf 43}, 3907 (1991)}

\maketitle

\thispagestyle{empty}



\begin{abstract}
Newton's equations for the motion of $N$ non-relativistic point particles
attracting according to the inverse square law may be cast in the form of
equations for null geodesics in a $(3N+2)$-dimensional Lorentzian spacetime
which is Ricci-flat and admits a covariantly constant null vector. 
Such a spacetime admits a Bargmann structure and corresponds physically to a
plane-fronted gravitational wave (generalized pp-wave).
Bargmann electromagnetism in five dimensions actually comprises the two distinct
Galilean electro-magnetic theories pointed out by Le Bellac and L\'evy-Leblond.
At the quantum level, the $N$-body Schr\"odinger equation may be
cast into the form of a massless wave equation. We exploit the conformal
symmetries of such spacetimes to discuss some properties of the Newtonian
$N$-body problem, in particular, (i) homographic solutions, (ii) the virial
theorem, (iii) Kepler's third law, (iv) the Lagrange-Laplace-Runge-Lenz vector
arising from three conformal Killing $2$-tensors and (v) the motion under
time-dependent inverse square law forces whose strength varies inversely as time
in a manner originally envisaged by Dirac in his theory of a time-dependent
gravitational constant $G(t)$. It is found that the problem can be reduced to
one with time independent inverse square law forces for a rescaled position
vector and a new time variable. This transformation (Vinti and Lynden-Bell) is
shown to arise from a particular conformal transformation of spacetime which
preserves the Ricci-flat condition originally pointed out by Brinkmann. We also
point out (vi) a Ricci-flat metric representing a system of $N$ non-relativistic
gravitational dyons. Our results for general time-dependent $G(t)$ are also
applicable by suitable reinterpretation to the motion of point particles in an
expanding universe. Finally we extend these results to the quantum regime.
\end{abstract}




\newpage
\section{Introduction}

Over the past few years, a new formalism has been developped for discussing the
symmetries of Galilei-invariant classical and quantum mechanical
systems \cite{LL1,JMS} associated with the non-relativistic spacetime
picture \cite{DBKP,Pru}.

The key point is that everything can be best handled on an {\em extended
spacetime}, a well behaved Lorentz manifold devised to tackle non-relativistic
physics. The situation is reminiscent of Kaluza-Klein theory : the classical
motions of a particle in a Newtonian potential correspond to {\em null geodesics}
in a $5$-dimensional ---or $(3N + 2)$-dimensional for $N$ particles--- Lorentz
manifold $(Q,g)$ carrying a covariantly constant null vector field $\xi$. These
so-called ``Bargmann structures'' were introduced in Ref.~\cite{DBKP}. 

The particle trajectories can also be obtained as the projections of {\em string}
trajectories in the extended spacetime. These strings moreover satisfy the
Polyakov equations of motion.

The null-geodesic/string formalism on our extended space\-time naturally leads
to studying the associated {\em conformal transformations} \cite{Jac0} that actually provide the chrono\-projec\-tive \cite{PBD}, Schr\"odinger \cite{Nie,Duv2},
Bargmann \cite{LL1,DBKP} and Galilei groups with a clear\-cut geometrical status. 

Conformally related Bargmann structures have the same null geodesics. We show
that the metric associated to a time-varying gravitational constant $G(t)$ is
conformally related to the $G_0$ case if and only if $G(t)$ changes according
to the prescription of Vinti \cite{Vin}, whose particular case is Dirac's
suggestion \cite{Dir}
\begin{equation}
G(t) = G_0\,{t_0 \over t}.
\label{1.1}
\end{equation}
	
The theory is readily extended to $N$ particles. The {\em generalized
Kepler's third law} and the {\em virial theorem} arise due to a certain
homothety of the corresponding metric.
The celebrated {\em Lagrange-Laplace-Runge-Lenz vector} of planetary motion
turns out to be associated with three conformal-Killing tensors of the
extended spacetime metric.

Incorporating full electromagnetism would necessitate an entirely relativistic
framework.
Electric and magnetic interactions can, however, be partially incorporated,
namely in two distinct ways. In one way one gets a ``magnetic'' theory
without the displacement current and in the other way one gets another
``electric'' theory where the Faraday induction term is missing. We here
recover by purely geometric means the two distinct theories of Galilean
electromagnetism originally discovered by Le Bellac and
L\'evy-Leblond \cite{LBLL}.

The Bargmann structures are the five-dimensional generalizations of the so 
called {\em pp-wave} solutions of Einstein's equations in four dimensions. The
latter describe {\em plane-fronted gravitational waves} and were discovered by
Brinkmann \cite{Bri}. They also admit a Kaluza-Klein interpretation allowing for
{\em magnetic mass} in dimension $D \ge 5$ and for a {\em Chern-Simons} type
modification of the field equations of pre-relativistic electromagnetism,
analogous to the situation studied recently by Jackiw \cite{Jac2}.

It is worth mentioning that Bargmann structures (in particular their pp-wave
solutions in arbitrary dimension) provide a wide class of time-dependent
classical solutions to string theory that have been extensively used \cite{HS}
to study string singularities.

In the case of the quantum $N$-body problem with the Dirac form (\ref{1.1}) for
$G(t)$, one is able (by using De Witt's quantization rules in curved space)
to associate to any solution of the Schr\"odinger equation for a time
independent Newtonian constant $G_0$, a solution of the Schr\"odinger equation
corresponding to a variable ``constant'' of gravitation $G(t)$. In the free
case one gets the quantum operators of the Schr\"odinger group, as written by
Niederer \cite{Nie}. It should be emphasized that, in our formalism,
quantization of these systems (and the appearance of symmetries) relies heavily
on the conformal invariance property of Schr\"odinger's equation in its
``Bargmannian'' guise.

Our present interest in non-relativistic conformal structures arose from
a study of Dirac's attempt to solve what is now known as the Hierarchy Problem.
Fifty years ago Dirac \cite{Dir} proposed in fact that Newton's gravitational
constant, $G$, varies inversely as the age, $t$, of the Universe (\ref{1.1}).
Thus $N$ celestial bodies of mass $m_j$ and position vectors ${\bf q}_j$ moving
non-relativistically would satisfy the equation :
\begin{equation}
{d^2 {\bf q}_j \over dt^2} = \sum_{k \ne j} G(t) m_k
{{\bf q}_k - {\bf q}_j \over \| {\bf q}_j - {\bf q}_k \|^3}
\qquad\hbox{where}\qquad
G(t) = G_0 {t_0 \over t}.
\label{1.2}
\end{equation}

An identical equation would arise if one supposed that the fine structure
constant varied inversely as the age of the universe and one considered the
classical  motion of nuclei and electrons in atoms and molecules. Of course in
that case it is more appropriate to consider the non-relativistic many particle
Schr\"odinger equation
\begin{equation}
i \hbar {\partial\Psi \over \partial t} = - {\hbar^2 \over 2}
\sum_{j=1}^N {1 \over m_j} \Delta_j \Psi + V \Psi
\qquad\hbox{where}\qquad
V = -\sum_{j<k} G(t) {m_j m_k \over \| {\bf q}_j - {\bf q}_k \|}.
\label{1.3}
\end{equation}

The classical equation (\ref{1.2}) also arises if one considers the motion of
particles in an  expanding  Universe with  scale factor $a$ in a conventional
theory in which Newton's constant really is constant $G=G_0$. If $T$ is cosmic
time then many people \cite{Pee} have pointed out that the relevant equation is
the cosmic Newton equation :
\begin{equation}
a^2 {d \over dT} \left(a^2 {d{\bf q}_j \over dT} \right)
= a(T) G_0 \sum_{k \ne j}{{\bf q}_k - {\bf q}_j \over
\| {\bf q}_j - {\bf q}_k \| ^3}.
\label{1.4}
\end{equation}
Equation (\ref{1.4}) can easily be cast in the form of equation (\ref{1.2}) with
an apparently time-dependent gravitational constant by defining a quasi-Newtonian
time coordinate $t$ by
\begin{equation}
dt = {dT \over a(T)^2}
\qquad\hbox{where}\qquad
G(t) = a(T) G_0.
\label{1.5}
\end{equation}

In order to obtain an apparent variation inversely as the Newtonian time
$t$ the scale factor $a(T)$ must be proportional to the cosmic time $T$.
This corresponds to an empty Milne model.

As emphasized by Lynden-Bell \cite{Lyn}, the classical equations (\ref{1.2}) are
no more or indeed no less difficult to solve than the usual equations with
constant coupling constant. That is, suppose we are given a solution
${\bf q}^*_j(t^*)$ of the classical equation of motion (\ref{1.2}) with $G=G_0$
independent of $t$, then
\begin{equation}
{\bf q}_j(t) = {t \over t_0}\ {\bf q}^*_j(-{t_0^2 \over t})
\label{1.6}
\end{equation}
is a solution of the equations of motion with time-dependent Newton's constant
varying inversely as the Newtonian time.

The corresponding statement for the quantum mechanical problem is : suppose that
we are given a solution $\Psi_{\rm static}({\bf q}_j^*,t^*)$ of the Schr\"odinger
equation (\ref{1.2}) with a time independent potential, then
\begin{equation}
\Psi({\bf q}_j,t) = \Big({t \over t_0}\Big)^{-3N/2}
\exp\Bigl({i \over 2 \hbar t} \sum_{j=1}^N m_j {\bf q}_j^2 \Bigr)
\Psi_{\rm static}(- {\bf q}_j {t_0 \over t},-{t_0^2 \over t})
\label{1.7}
\end{equation}
is a solution of the Schr\"odinger equation with a fine structure constant
varying inversely as the Newtonian time.
The pre-factor $(t/t_0)^{-3N/2}$ guarantees that the wave function remains
normalised.

These two results are exact and may readily be verified by elementary
differentiations.

Dirac's original hypothesis is excluded by observations, however for times
short compared with $t_0$ any time variability of $G$ can be modeled by a
$1/t$-law, and these classical results are useful in discussing observations of
the binary pulsar \cite{DGT,DGG}.

These results provide a fascinating  example where both the classical and the
quantum mechanical adiabatic theorems are in some sense exact. Their truth is
suggested by the work of Vinti \cite{Vin} who pointed out in the context of the
two body problem in Dirac's theory, the relevance of a result of
Mestschersky \cite{Mes} showing that the $2$-body problem could be reduced to
quadratures. The reader is referred to Vinti's paper for a detailed
consideration of the solutions in the case $N=2$.

\goodbreak

In order to explain the relation between the situations with
$G(t)$ and $G_0$, remember that the equation of motion (\ref{1.2}) may be
derived from the action
\[
S = \int\!\left(\sum_{j=1}^N{m_j \over 2}\Big({d{\bf q}_j \over dt}\Big)^2
- {1 \over 2}\sum_{j \ne k} G(t) 
{m_j m_k \over \| {\bf q}_j - {\bf q}_k \|}\right)\,dt.
\]
Consider furthermore the transformation \cite{Vin,Lyn}
\begin{equation}
D : ({\bf q}_j, t) \to ({\bf q}_j^*,t^*) = 
({t_0 \over t} {\bf q}_j,-{t_0^2 \over t})
\label{1.9}
\end{equation}
where $j = 1,\ldots, N$. It is easy to see that if $G(t)$ changes as suggested
by Dirac (\ref{1.1}), the action, $S$, changes by a mere boundary term,
\begin{equation}
S = S^* - \int\! d\left(\sum_{j=1}^N {m_j {{\bf q}_j^*}^2 \over 2t^*}\right),
\label{1.10}
\end{equation}
where $S^*$ is the action with $G_0$. This explains the result of Lynden-Bell.

Observe that the above total derivative actually comes from the kinetic term.
The ``Vinti-Lynden-Bell transformation'', $D$ in (\ref{1.9}), is therefore 
a symmetry for a free particle. But the symmetries of a free particle form 
a $2$-parameter extension of the Galilei group, the so-called
``Schr\"odinger group'' \cite{Nie,Duv2,HSHJ}. 
The new
transformations are
\begin{equation}
\delta_d : ({\bf q},t) \to (d{\bf q},d^2t),
\qquad
d\in{\bf R}\backslash\{0\}
\qquad 
\hbox{(dilatations)},
\label{1.11}
\end{equation}
\begin{equation}
\kappa_k : ({\bf q},t) \to (\frac{{\bf q}}{1 - kt},\frac{t}{1 - kt}),
\qquad  k\in{\bf R} \qquad \hbox{(expansions)}.
\label{1.12}
\end{equation}
They form, together with
\begin{equation}
\epsilon_e : ({\bf q},t) \to ({\bf q},t + e), \qquad  e\in{\bf R}
\qquad  \hbox{(time translations)},
\label{1.13} 
\end{equation}
a group isomorphic to $SL(2,{\bf R})$. This group was later rediscovered \cite{Jac1}
as a symmetry of the Dirac monopole and more recently \cite{Jac2} as a symmetry
of a magnetic vortex.

Being a symmetry of a free particle, the Vinti-Lynden-Bell transformation
(\ref{1.9}) is expected to belong to the Schr\"odinger group. It is easy to see
that this is indeed the case since, for $t_0 \ne 0$, we have
\[
D =
\epsilon_{t_0}
\,{\raise 0.5pt \hbox{$\scriptstyle\circ$}}\,\kappa_{1/t_0}
\,{\raise 0.5pt \hbox{$\scriptstyle\circ$}}\,\epsilon_{t_0}
\,{\raise 0.5pt \hbox{$\scriptstyle\circ$}}\,\delta_{-1}.
\]

In this paper we deal exclusively with the case of $3$ spatial dimensions.
However, all of our considerations generalize in a straightforward fashion to
an arbitrary number, $d$, of spatial dimensions. In this case, the analogue of
Dirac's law for the time dependence of $G$ is
$$
G(t) \propto {1\over t^{d - 2}}.
$$

\goodbreak

The organization of the paper is as follows.

-- In Sec.~2 we provide a review of the geometrical framework we use, that
is  of Bargmann structures leading to a covariant formulation of classical
mechanics and pre-relativistic electromagnetism.

-- In Sec.~3 we discuss the single particle case with varying ``constant'' 
of gravitation.

-- In Sec.~4 we review the conformal symmetries of a Bargmann structure and
relate these to the so-called Vinti-Lynden-Bell transformations.

-- Sec.~5 extends this work to the $N$-body problem by passing to a $(3N +
2)$-dimensional spacetime and deals with the homographic solutions and the
virial theorem.

-- In Sec.~6 we show how the Lagrange-Laplace-Runge-Lenz vector is associated
to a 3-vector's worth of conformal Killing $2$-tensors.

-- In Sec.~7 we relate our work to that of Brinkmann and we also point out that
Newtonian theory allows for a natural generalization including
gravitational ``magnetic'' mass monopoles.

-- Finally, in Sec.~8 we show how our results can be extended to the quantum
regime, in particular how mass gets quantized in the presence of gravitational
dyons in this non-relativistic context.


\section{Bargmann structures and covariant mechanics}

In this section we give a short outline of a geometrical framework adapted to
the description of non-relativistic classical and quantum physics.

\subsection{Bargmann structures}

Let us first recall that a Bargmann manifold \cite{DBKP} is a
$5$-dimensional Lorentz manifold $(Q,g)$ with signature $(++++-)$  and a
fibration by a free action of the additive group of real numbers, $({\bf R},+)$,
whose infinitesimal generator $\xi$ is null and covariant constant.

A $SO(2)$-action would lead to a topologically non trivial bundle suitable to
generalize Newtonian theory as described in Secs 7 and 8.
 
As an example, let us consider the extended spacetime
$({\bf R}^3 \times {\bf R}) \times {\bf R} \ni ({\bf q},t,s)$, where the fifth coordinate $s$
has dimension [action]/[mass]; start off with the following Lorentz
metric
\begin{equation}
g_U \equiv \langle d{\bf q} \otimes d{\bf q} \rangle + dt \otimes ds +
ds \otimes dt - 2U({\bf q},t)\,dt\otimes dt
\label{2.1}
\end{equation}
where $\langle\ ,\ \rangle$ is the standard Euclidean scalar product on ${\bf R}^3$
and put
\begin{equation}
\xi \equiv {\partial\over\partial s}.
\label{2.2}
\end{equation}
It is a simple task to check that $({\bf R}^5,g_U,\xi)$ is actually a Bargmann
manifold.

The base manifold $B = Q/({\bf R},+)$ is readily interpreted as spacetime. If we
denote by $\pi : Q \to B$ the corresponding projection, we see that $B$ is
canonically endowed with a Galilei structure, i.e. a symmetric $2$-contravariant
tensor $h = \pi_*g^{-1}$ with signature $(+++\,0)$ and a $1$-form $\theta$
defined by $g(\xi) = \pi^* \theta$ which generates $\ker(h)$. Note that the
``clock'' $\theta$ is closed and the integrable space-like distribution
$\Sigma = \ker(\theta)$ then defines the {\em absolute time} axis $B/\Sigma$.

In our example (\ref{2.1},\ref{2.2}), we easily find that spacetime 
$B \cong {\bf R}^3 \times {\bf R}$
has the following Galilei structure :
$h = \langle{\partial/\partial{\bf q}}\otimes{\partial/\partial{\bf q}}\rangle,
\theta = dt$.

The Levi-Civita connection $\nabla^{(Q)}$ on $Q$ can be shown to descend
to $B$ as a preferred symmetric ``Newton-Cartan'' connection, \cite{Car,Kun}
$\nabla^{(B)}$, (which, in particular, parallel-transports $h$ and $\theta$)
interpreted as the Newtonian gravitational field. In the particular case
(\ref{2.1},\ref{2.2}), one finds, using obvious  notations,
$\Gamma_{tt}^A = \partial_A U = -\Gamma_{At}^s \ (A = 1,2,3)$ and 
$\Gamma_{tt}^s = -\partial_t U$ for the non-zero Christoffel symbols of
$\nabla^{(Q)}$.
A straightforward computation shows furthermore that the geodesics of $(Q,g_U)$
project upon spacetime $(B,h,\theta,\nabla^{(B)})$ as the worldlines of test
particles in the gravitational potential $U$.

Consider now such a geodesic $(\tau \to q(\tau))$ and set $p = dq/d\tau$. 
The two quantities $p^2 \equiv g_{ab} p^a p^b$ and $m \equiv g_{ab} p^a \xi^b$
($a,b = 1,\ldots,5)$ are conserved along the motion. Comparing with the free
case, the $3$-(co)vector ${\bf p} = (p_1\,p_2\,p_3)$ may be interpreted as the
linear momentum, $-p_t = E$ as the energy and $p_s = m$ as the mass.
An easy calculation yields
$$
p^2 = 2m\,\left({{\bf p}^2 \over 2m} - E + V\right) = - 2mE_0,
$$
where $V \equiv m U$ and $E_0$ is thus the {\em internal energy} of the particle
under consideration. But, in the non-relativistic theory, the internal energy is
required to vanish, $E_0 = 0$. The motion of our test particle of mass $m$ in the
potential $V$ can be described therefore by a {\em null geodesic} in the
extended spacetime. We defer to Sec.~5 the general case of $N$ bodies.

Let us now assume that two Bargmann structures $(Q,g,\xi)$ and $(Q,g^*,\xi^*)$
on the same spacetime extension are related by 
\begin{equation}
g^* = \Omega^2 g \qquad\hbox{and}\qquad \xi^* = \xi.
\label{2.3}
\end{equation} 
The nowhere vanishing function $\Omega$ is necessarily a function of time only
since the new clock $\theta^* = \Omega^2 \theta$ must be closed, implying
$d\Omega \wedge dt = 0$. But conformally related metrics have the same null
geodesics, yielding the same world lines in spacetime. Due to (\ref{2.3}), the
mass is also preserved.

\goodbreak

\subsection{Symplectic mechanics \& strings}


Classical mechanics can be most elegantly formulated in a symp\-lectic
formalism \cite{JMS}. We thus wish to present, in this geometric setup, the
equations of motion of a point particle dwelling in spacetime and subject to
the action of an external gravitational field described by a Bargmann structure.
	
We start with the cotangent bundle $T^*Q$ endowed with its canonical $1$-form
$\vartheta \equiv p_a dq^a$. The equations of motion we are interested in can be
obtained as follows. Consider the (closed) $8$-dimensional submanifold  $C
\hookrightarrow T^*Q$ defined  by the two previously introduced  constraints
$(a,b = 1,\ldots,5)$ : \begin{equation}
E_0 = g^{ab} p_a p_b = 0 
\label{2.4}
\end{equation}
and
\begin{equation}
m \equiv p_a \xi^a = \mathop{\rm const}\nolimits.
\label{2.5}
\end{equation}
We will confine considerations to massive systems only.
The restriction $\omega_C = d\vartheta\vert_C$ to $C$ of the symplectic $2$-form
of $T^*Q$ turns $C$ into a presymplectic manifold with a $2$-dimensional null
foliation $\ker(\omega_C)$, whose leaves  projects onto $Q$ as $2$-surfaces,
physically the world sheets of {\em strings}. These project in turn onto
spacetime $B$ as curves, i.e. as the world-lines of {\em particles}. The
equations of the foliation are indeed
\begin{equation}
\left\{
\begin{array}{ll}
\dot p_a = 0 & \\
\dot q^a = \alpha\,p^a + \beta\,\xi^a, \qquad & (\alpha,\beta\in{\bf R}),
\end{array}
\right.
\label{2.6}
\end{equation}
where the dot in the first equation stands for covariant derivative with respect
to $\nabla^{(Q)}$. We thus have the diagram
$$
\begin{array}{ccc}
T^*Q &
\smash{\mathop{\hbox to 18mm{\leftarrowfill}}
\limits^{\scriptstyle{E_0 = 0}}_{\scriptstyle{m = {\rm const}}}}
\!\!\!\rhook & C \\
\llap{$\scriptstyle{}$}\left\downarrow
\vbox to 7mm{}\right.\rlap{$\scriptstyle{}$}
& &
\llap{$\scriptstyle{{\rm ker}(\omega_C)}$}\left\downarrow
\vbox to 7mm{}\right.\rlap{$\scriptstyle{{\bf R}^2}$}
\\
Q & & X
\end{array}
$$
where $X = C/\ker (\omega_C)$ is a $6$-dimensional manifold canonically
endowed with the symplectic $2$-form $\omega_X$, the projection of $\omega_C$.
In Souriau's terminology, \cite{JMS} $(X,\omega_X)$ is the {\em space of motions}
of our test particle.

Hence the geodesics of $(B,\nabla^{(B)})$ arise from strings in the
Bargmann manifold.
Interestingly enough, these strings satisfy the {\em Polyakov equations of
motion} in $(Q,g)$. (An analogous situation appears for ``winders'' in
$5$-dimensional Kaluza-Klein theory, \cite{GR} where the strings satisfy the
Goto-Nambu equations.) To see this, note that the Polyakov action for a string,
i.e. for a map
$
q^a :
\left(M_2,(\gamma_{ij})\right) \to \left(Q,(g_{ab})\right) :
(\sigma^1,\sigma^2) \to q^a(\sigma^1,\sigma^2)
$
, is given by \cite{GSW}
\begin{equation}
\int\!{g_{ab}\,
{\partial q^a \over \partial\sigma^i}
{\partial q^b \over \partial\sigma^j}
\,\gamma^{ij}
\sqrt{\gamma}\,d\sigma^1 d\sigma^2}.
\label{2.7}
\end{equation}
Varying with respect to the $q$'s gives
\begin{equation}
\Delta _\gamma q^a + \Gamma^a_{bc}\,{\partial_i q^b}{\partial_j q^c}\,
\gamma^{ij} = 0,
\label{2.8}
\end{equation}
and varying with respect to the $\gamma$'s gives
\begin{equation}
g_{ab}\left(
{\partial_i q^a}{\partial_j q^b}
-\mbox{$\textstyle{1\over 2}$}
\gamma_{ij}\gamma^{mn}{\partial_m q^a}{\partial_n q^b}
\right) = 0.
\label{2.9}
\end{equation}
In the present case, we set
$\gamma =
\gamma_{ij}\,d\sigma^i \otimes d\sigma^j = m(du \otimes dv + dv \otimes du)$
for the $2$-metric and these equations become
\begin{equation}
{\partial^2 q^a \over \partial u \partial v}
+ \Gamma^{a}_{bc} 
{\partial q^b \over \partial u}{\partial q^c \over \partial v} = 0,
\label{2.10}
\end{equation}
and
\begin{equation}
g_{ab} \partial_u q^a \partial_u q^b = 
g_{ab} \partial_v q^a \partial_v q^b = 0.
\label{2.11}
\end{equation}
The previously introduced coordinate system $(u,v)$ on $M_2$ is actually
given by
\begin{equation}
p^a \equiv {\partial q^a \over \partial u}
\qquad\hbox{and}\qquad
\xi^a \equiv {\partial q^a \over \partial v},
\label{2.12}
\end{equation}
and Eq. (\ref{2.10}) is thus equivalent to $\nabla_\xi\,p^a = \nabla_p\,\xi^a =
0$, a system which is clearly satisfied by our string-equations of motion
(\ref{2.6}) ---the second equation being automatic since $\nabla\xi \equiv 0$ on
a Bargmann manifold. On the other hand, the constraint (\ref{2.4}) and the
isotropy condition $g(\xi,\xi) = 0$ become just the two equations in
(\ref{2.11}).

\subsection{Galilean electromagnetisms}


We conclude this section with a novel viewpoint on the two distinct theories of
Galilean electromagnetism.

\subsubsection{The magnetic-like theory}

Consider firstly a $2$-form
${\cal F} = {1\over 2} {\cal F}_{ab}\,dq^a \wedge dq^b$ on the extended
(Bargmann) spacetime which satisfies the $5$-dimensional Maxwell equations
$(a,b,c = 1,\ldots,5)$ :
\begin{equation}
\partial_{[a} {\cal F}_{bc]} = 0,
\label{2.13}
\end{equation}
\begin{equation}
\nabla^a {\cal F}_{ab} = {\cal J}_b.
\label{2.14}
\end{equation}
(Square brackets denote skew-symmetrization.)
	
In order to get a well behaved $4$-dimensional theory, we require furthermore
that
\begin{equation}
{\cal F}_{ab} \xi^a = 0.
\label{2.15}
\end{equation}
Now, ${\cal F}$ being closed, this last condition entails that
$L_\xi{\cal F} = 0$, hence that ${\cal F}$ is the pull-back of a $2$-form
$F = \frac{1}{2} F_{\alpha\beta}\,dq^\alpha \wedge dq^\beta$
on spacetime $B$.
We clearly still have $dF = 0$, i.e. the first group of Maxwell's equations,
\begin{equation}
\partial_{[\alpha} F_{\beta\gamma]} = 0
\label{2.16}
\end{equation}
where $\alpha,\beta,\gamma = 1,\ldots,4$.

We then reduce the second group (\ref{2.14}). Since $\xi$ is covariantly
constant, Eq. (\ref{2.15}) readily implies that the $5$-current ${\cal J}$ has no
``fifth component'' (i.e. ${\cal J}_a \xi^a = 0$), or, componentwise
$({\cal J}_a) = ((J_\alpha),0)$.
The only non-vanishing Christoffel symbols of $\nabla^{(Q)}$ are
$\Gamma^\alpha_{\beta\gamma}$ and $\Gamma^5_{\alpha\beta}$ in an adapted
coordinate system. The explicit form of $\Gamma^5_{\alpha\beta}$ is given by Eq.
(3.25) in Ref.~\cite{DBKP} and will not be needed here.
The $\Gamma^\alpha_{\beta\gamma}$'s turn out to be nothing
but the components of the Newtonian connection $\nabla^{(B)}$ on spacetime. The
$5$-dimensional equation (\ref{2.14}) reduces therefore to
\begin{equation}
h^{\alpha\beta} \nabla_\alpha F_{\beta\gamma} = J_\gamma
\label{2.17}
\end{equation}
where $(h^{\alpha\beta})$ is the $4$-dimensional Galilei ``metric''.
Had we worked in a truly relativistic spacetime
$\left(B,(g^{\alpha\beta})\right)$ by considering, from the outset, a
spacelike fibration \cite{DBKP,KD} $g(\xi,\xi) = c^{-2}$, Eq.~(\ref{2.17})
would have yielded the second group of the Maxwell equations. However, our
``metric'' $h$ is degenerate; as a result {\em the displacement current is
lost} here. For example, in flat spacetime
$\left({\bf R}^4,
(h^{\alpha\beta}) = \mathop{\rm diag}\nolimits(1,1,1,0),
(\theta_\alpha) = (0\,0\,0\,1)\right)$,
Eqs (\ref{2.16},\ref{2.17}) reduce to
\begin{equation}
\mathop{\rm div}\nolimits\,{\bf B} = 0,\qquad \mathop{\bf curl}\nolimits
\,{\bf E} + {\partial{\bf B}\over\partial t}
= {\bf 0},
\label{2.18}
\end{equation}
\begin{equation}
\mathop{\rm div}\nolimits\,{\bf E} = \rho,\qquad \mathop{\bf curl}\nolimits
\,{\bf B} = {\bf j}.
\label{2.19}
\end{equation}
We have put $E^A \equiv F_{A4},\,B^C \equiv {1\over 2} \epsilon^{ABC}F_{AB}$
(with $A,B,C = 1,2,3$) to define the Galilean electromagnetic field
$({\bf E},{\bf B})$ and $j^A \equiv J_A,\,\rho \equiv J_4$ to define the
$3$-current density $\bf j$ and the charge density $\rho$.

In the curved case, however, Eqs (\ref{2.16},\ref{2.17}) retain the same form as
in  the flat case, except for a possible modification of Gauss' law,
\begin{equation}
\mathop{\rm div}\nolimits\,{\bf E} + \langle {\bf H},{\bf B} \rangle = \rho
\label{2.20}
\end{equation}
that couples the electromagnetic field to the ``Coriolis''
components $H^C \equiv \epsilon_A^{\ BC} \Gamma^A_{B4}$  (see
(\ref{7.20})) of the gravitational field corresponding to certain
solutions of Newton-Cartan field equations, such as the non-relativistic
Taub-NUT type solutions given by the metric (\ref{7.7}). This modification is
very reminiscent of the Chern-Simons modification of Maxwell's
equations \cite{Jac2}. This point will not be pursued here.

This is the so-called ``magnetic limit'' of Maxwell's equations.

\subsubsection{The electric-like theory}

Interestingly enough, the $5$-dimensional Maxwell equations admit another
subtle ``electric limit'' \cite{LBLL,Kun2}. We could have started with 
the {\em contra\-variant} form of the $5$-dimensional Maxwell 
equations, i.e. with
\[
\partial^{[a} {\cal F}^{bc]} = 0,
\qquad\hbox{and}\qquad
\nabla_a  {\cal F}^{ab} = {\cal J}^b,
\]
which is obviously equivalent to the covariant form (\ref{2.13},\ref{2.14}). 
By requiring now the mere invariance condition $L_\xi{\cal F} =0$, one can
push-down the $2$-vector ${\cal F}^{ab}\,\partial_a \otimes \partial_b$ we still
denote ${\cal F}$ and get a $2$-vector $\widetilde F = \pi_*{\cal F}$ on
spacetime $B$ that satisfies $(\alpha,\beta,\gamma,\ldots = 1,\ldots,4)$ :
\begin{equation}
\nabla_\alpha \widetilde F^{\alpha\beta} = \widetilde J^\beta,
\label{2.22}
\end{equation}
where $\widetilde J = \pi_*{\cal J}$ and
\begin{equation}
h^{\lambda [\alpha} \partial_\lambda \widetilde F^{\beta\gamma]} = 0,
\label{2.23}
\end{equation}
that is, the {\em full second group} of Maxwell's equations with the
displacement current included, which retains the following familiar form in
flat spacetime
\begin{equation}
\mathop{\rm div}\nolimits\,\widetilde{\bf E} = \widetilde\rho,
\qquad
\mathop{\bf curl}\nolimits\,{\bf B} - 
{\partial\widetilde{\bf E} \over \partial t} = {\bf j},
\label{2.24}
\end{equation}
together with a {\em truncated first group}, i.e. with the induction term
missing
\begin{equation}
\mathop{\rm div}\nolimits\,{\bf B} = 0,\qquad \mathop{\bf curl}\nolimits
\,\widetilde{\bf E} = {\bf 0}.
\label{2.25}
\end{equation}
Notice that we have $\widetilde{\bf B} = {\bf B}$ if we again define the
``electric-like'' electromagnetic field
$(\widetilde{\bf E},\widetilde{\bf B})$
by
$\widetilde E^A \equiv \widetilde F^{A4},\,
\widetilde B^C \equiv {1\over 2} \epsilon_{ABC}\widetilde F^{AB}$
and the current-charge density by
$\widetilde j^A \equiv\widetilde J^A = j^A,\,
\widetilde\rho \equiv\widetilde J^4$.

In addition, there are extra fields
$E^A \equiv {\cal F}^{A5},\,S \equiv {\cal F}^{45}$ which push-forward to zero.
One easily finds that they satisfy in the flat case :
$\mathop{\bf grad}\nolimits\,S + \partial\widetilde{\bf E}/\partial t = {\bf 0},
\, \mathop{\bf curl}\nolimits\,{\bf E} + \partial{\bf B}/\partial t = {\bf 0}$
and
$\mathop{\rm div}\nolimits\,{\bf E} + \partial S/\partial t = \rho$
where $\rho \equiv {\cal J}^5$. 
By contrast with the more familiar Kaluza-Klein case when the fibration is
spacelike, one cannot, in our case where the fibration is null,
associate in an intrinsic fashion these extra components to fields on
spacetime.


\section{A time varying ``constant'' of gravitation}

We now apply our general framework to cope with Dirac's theory in which the
gravitational ``constant'' varies with the time. 

On a Bargmann manifold $(Q,g,\xi)$ Newton's field equations for gravity can be
cast into the simple geometrical form \cite{DBKP}
\begin{equation}
R_{ab} = 4\pi G\varrho\,\xi_a \xi_b
\label{3.1}
\end{equation}
where $\varrho$ is the mass density of the sources. The spacetime projection
of these equations yields the so-called Newton-Cartan field equations \cite{Kun}.
Those reduce, in the particular case of a Bargmann manifold $(Q,g_U,\xi)$ given
by (\ref{2.1},\ref{2.2}) with $Q \subset {\bf R}^5$, to the {\em Poisson equation}
$$
\Delta U = 4\pi G\varrho.
$$

Since the gravitational ``constant'', $G$, can be consistently assumed to
depend on time only,
\begin{equation}
U({\bf q},t) \equiv - G(t)\,{m_0 \over \|{\bf q}\|}
\qquad\hbox{with}\qquad
G_0 \equiv G(t_0)
\label{3.2}
\end{equation}
plainly defines a $SO(3)$-invariant vacuum solution of (\ref{3.1}) on 
$Q = ({\bf R}^3 \backslash \{0\}\times{\bf R})\times{\bf R}$.

By contrast with Einstein's theory in which the constant of gravitation must be
independent of time, unless one augments the theory by additing an extra
scalar field, \cite{Jor,BD} a time-dependent gravitational ``constant'' is
quite natural in the Newtonian context. In fact, even if one is interested in
the usual case where $G$ is constant with time, our results on time-dependent
$G(t)$ are directly applicable to the motion of particles in an expanding
universe (Eq. (\ref{1.4}) using the identification (\ref{1.5})).

Suppose now that we can find a (local) diffeomorphism $\cal D$ of $Q$ with
the metric $g_U$ given by (\ref{2.1}) and (\ref{3.2}) and the vertical vector
$\xi$ given by (\ref{2.2}) such that
\begin{equation}
{\cal D}^*g_U = \Omega^2\,g_{U_0}
\label{3.3}
\end{equation}
for some strictly positive function $\Omega$ and
\begin{equation}
{\cal D}_*\xi = \xi,
\label{3.4}
\end{equation}
where $U_0({\bf q}) = U({\bf q},t_0)$. The metrics ${\cal D}^*g_U$ and $g_{U_0}$
clearly have the same null geodesics, whence the same world lines in spacetime.
Accordingly, up to a change of coordinates, the equations of motion governed by
$G(t)$ are the same as those corresponding to $G_0$. Let us show that this
happens for very special functions $G(t)$.

We seek the conformal equivalence (\ref{3.3}) by putting
$({\bf q}^*,t^*,s^*) \equiv {\cal D}({\bf q},t,s)$
as  a shorthand.
Being a bundle automorphism (\ref{3.4}), ${\cal D}$ defines a spacetime local
diffeomorphism $({\bf q},t) \to ({\bf q}^*,t^*)$ which, in turn, defines a local
diffeomorphism $t \to t^*$ of the time axis.
Owing to the fundamental $SO(3)$-symmetry of the problem, we confine
considerations to $SO(3)$-equivariant diffeomorphisms, i.e.
${\bf q}^* = \Omega (r,t)\,{\bf q}$ where $r \equiv \|{\bf q}\|$.
But, as previously noticed (\ref{2.3}), $\Omega$ cannot depend on $r$; after 
some calculation we get $t^* = \int\!\Omega(t)^2\,dt$ and
\begin{eqnarray*}
{\cal D}^*g_U = &\Omega(t)^2& \left(\langle d{\bf q} \otimes d{\bf q} \rangle
+ dt \otimes ds + ds \otimes dt
+ 2\,{G_0 m_0 \over r}\,dt \otimes dt \right)\\
+ &\Omega(t)^2& \left(dt \otimes \psi + \psi \otimes dt \right)
\end{eqnarray*}
with
$$
\psi = ds^* - ds + \alpha (t) r dr +
\mbox{$\textstyle{1\over 2}$}
\alpha (t)^2 r^2 dt +
2\,{\beta (t) m_0 \over r} dt
$$
where
$\alpha (t) \equiv {\dot\Omega (t)/\Omega (t)}$
and
$\beta (t) \equiv G(t^*) \Omega (t) - G_0$.
Now Eq. (\ref{3.3}) will be satisfied provided $\psi$ vanishes. But this
requires
$
d\psi = (\dot\alpha - \alpha^2 + 2 \beta/r^3)\,dt\wedge dr = 0,
$
that is $\dot\alpha = \alpha^2$ and $\beta = 0$.
Hence
$$
G(t^*) = {G_0 \over \Omega(t) }
\qquad\hbox{and}\qquad
\Omega(t) = {a \over t + b},
$$
with $a,b = \mathop{\rm const}\nolimits$.
We thus obtain
$s^* = s - {1 \over 2} \alpha (t) r^2 + d$ with $d = \mathop{\rm const}\nolimits$. Condition
(\ref{3.4}) then holds automatically because
\[
{\partial\over\partial s^*} = {\partial\over\partial s}.
\]
Thus $t^* = - a^2/(t + b) + c$ with $c = \mathop{\rm const}\nolimits$. Let us now summarize.

\begin{em}
The solution $g_U$ of Newton's field equation associated to
$$
U({\bf q},t) = - G(t)\,{m_0 \over \| {\bf q} \|}
$$
can be $SO(3)$-equivariantly brought into the conformal class of $g_{U_0}$ 
where
$$
U_0({\bf q}) = - G_0\,{m_0 \over \| {\bf q} \|},
$$
provided
\begin{equation}
G(t) = G_0\,{a \over t - c}.
\label{3.6}
\end{equation}
The local diffeomorphisms, $\cal D$, which commute with the rotations and such
that ${\cal D}^*g_U = \Omega^2 g_{U_0}$ and ${\cal D}_*\xi = \xi$ will be called
``Vinti-Lynden-Bell transformations''. They are given by
\begin{equation}
\left\{
\begin{array}{l}
{\bf q}^* = {\displaystyle {a \over t + b}}\,{\bf q}\\[6pt]
t^* = - {\displaystyle {a^2 \over t + b}} + c\\[6pt]
s^* = s + {\displaystyle {{\bf q}^2 \over 2(t + b)}} + d,
\end{array}
\right.
\label{3.7}
\end{equation} 
$a \ne 0,b,c$ and $d$ being arbitrary real constants.
The conformal factor is finally
\[
\Omega(t) = {a \over t + b}.
\]
\end{em}

This result is consistent with that of Ref.~\cite{Vin}. Dirac's original
suggestion (\ref{1.1}) for $G(t)$ corresponds to choosing $a = t_0, b = c = 0$ 
in (\ref{3.7}). Vinti actually proposed the slight modification (\ref{3.6}) to
Dirac's prescription in order ``to avoid infinities in the resulting exact
classical solutions for the orbit in the two-body problem''. Here, we get Vinti's
formula as the general solution of the problem of conformal equivalence for
Newtonian gravity between a constant and a time-varying ``constant'' of
gravitation.

Notice that the change (\ref{3.7}) in the extra coordinate $s$ is precisely the
change of the action as given in (\ref{1.10}) and that the spacetime part of
(\ref{3.7}) can actually be obtained by having the $5 \times 5$ matrices
\begin{equation}
\pmatrix{
{\bf 1} & {\bf 0} & {\bf 0} \cr
{\bf 0} & {c/a} & {bc/a - a} \cr 
{\bf 0} & {1/a} & {b/a} \cr
}
\label{3.9}
\end{equation}
act projectively on the affine space of the $5$-vectors
$$
\pmatrix{
{\bf q} \cr
t \cr
1 \cr
}
$$
representing spacetime events.
We record for further purposes that the matrices (\ref{3.9}) form an open subset
of $SL(2,{\bf R})$.


\section{Bargmann conformal symmetries}

In this section we discuss the general notion of conformal symmetries of the
$5$-dimensional Bargmann spacetime. It will be shown in Sec.~8 that these
symmetries are actually specific to the Schr\"odinger equation.

It was recognized by Niederer \cite{Nie} and Hagen \cite{Hag} in the early
seventies that the maximal kinematical symmetry group of the free Schr\"odinger
equation is larger than the mere Bargmann group, i.e. the $11$-dimensional
extended Galilei group. This group is $13$-dimensional and has been called the
``extended Schr\"odinger group''. In our formalism, it simply consists of
those conformal transformations $\cal C$ of the canonical flat Bargmann
structure $({\bf R}^5,g_0,\xi)$, the extended Galilei spacetime, that commute
with the structural group, i.e. such that
\begin{equation}
{\cal C}^*g_0 = \Omega^2 g_0
\qquad\hbox{and}\qquad
{\cal C}_*\xi = \xi,
\label{4.1}
\end{equation}
with
\begin{equation}
g_0 = \langle d{\bf q} \otimes d{\bf q} \rangle
+ dt \otimes ds + ds \otimes dt
\qquad\hbox{and}\qquad
\xi = {\partial \over \partial s}.
\label{4.2}
\end{equation}
See Refs~\cite{Duv2,PBD} for a more detailed account.
It is amusing to note that the conformal transformations were already used
in five dimensions to study the parabolic diffusion equation at the beginning
of the century \cite{Bat}.

Let us first determine those conformal transformations of
$({\bf R}^5,g_0,\xi)$ which simply project down to the base $B$ as spacetime
transformations, i.e. which preserve the vertical {\em direction}.
Infinitesimally, this amounts to finding all vector fields $X$ such that
\begin{equation}
L_X\,g_0 = 2\lambda\,g_0
\qquad\hbox{and}\qquad
L_X\,\xi = \mu\,\xi,
\label{4.3}
\end{equation}
for some functions $\lambda$ and $\mu$.
The solutions of this system form a $14$-dimensional Lie algebra for the Lie
bracket (the so-called ``chronoprojective Lie algebra'') and are given by
\begin{equation}
\left\{
\begin{array}{l}
{\bf X} = 
\mbox{\boldmath$\alpha$} \times {\bf q} + (\chi + \kappa t){\bf q}
+ \mbox{\boldmath$\beta$}t + \mbox{\boldmath$\gamma$}\\[4pt]
X^t = \kappa t^2 + \delta t + \epsilon\\[4pt]
X^s = - \left(
\mbox{$\textstyle{1\over 2}$}
\kappa {\bf q}^2
+ \langle \mbox{\boldmath$\beta$},{\bf q} \rangle
+ \eta + (\delta - 2\chi) s \right),
\end{array}
\right.
\label{4.4}
\end{equation}
with
$\mbox{\boldmath$\alpha$},\mbox{\boldmath$\beta$},\mbox{\boldmath$\gamma$}
\in{\bf R}^3;
\chi,\kappa,\delta,\epsilon,\eta\in{\bf R}$.
This yields $\lambda = \chi + \kappa t$ and $\mu = \delta - 2\chi$ and the
subalgebra of conformal bundle automorphisms $(\mu = 0)$ is thus characterized by
\begin{equation}
\chi = {\delta\over 2}.
\label{4.6}
\end{equation}

Integrating this Lie algebra leaves us with a $13$-dimensional Lie group, the
(neutral component of the) so-called {\em extended Schr\"odinger group}
``acting'' on the extended spacetime according to
\begin{equation}
\left\{
\begin{array}{l}
{\bf q}^* = {\displaystyle {A{\bf q} + {\bf b}t + {\bf c} \over ft + g}}\\[8pt]
t^* = {\displaystyle {dt + e \over ft + g}}\\[8pt]
s^* = {\displaystyle
s
+ \frac{f}{2}{\left(A{\bf q} + {\bf b}t + {\bf c}\right)^2 \over ft + g} 
- \langle{\bf b},A{\bf q}\rangle 
- \frac{t}{2}{\bf b}^2 + h,
}
\end{array}
\right.
\label{4.7}
\end{equation}
where $A \in SO(3); {\bf b},{\bf c} \in {\bf R}^3; d,e,f,g,h \in {\bf R}$ and
$dg - ef = 1$. The corresponding conformal factor in (\ref{4.1}) is therefore
\[
\Omega = {1 \over ft + g}.
\]

\goodbreak

It is now possible to give other non-relativistic ``symmetry'' 
groups (listed below) a neat geometrical interpretation associated with the
flat Bargmann structure.

i) The $11$-dimensional subgroup defined by $d = g = 1$, $f = 0$ in (\ref{4.7})
is the (neutral component of the) {\em Bargmann group} which consists of those
$\xi$-preserving {\em isometries} of the extended spacetime.

ii) The {\em Galilei group} is recovered as the $10$-dimensional quotient
of the Bargmann group by its centre, $({\bf R},+)$, parametrized by $h$ (see
(\ref{4.7})). The action of the restricted Galilei group on spacetime is
given by the first two equations in (\ref{4.7}) with $d = g =1$, $f = 0$; it
corresponds to the projection onto spacetime ${\bf R}^4$ of the Bargmann group
action on ${\bf R}^5$, the extended spacetime.

iii) Again, factoring the extended Schr\"odinger group (\ref{4.7}) by its centre,
$({\bf R},+)$, yields the $12$-dimensional {\em Schr\"odinger group}, originally
discovered as the ``maximal invariance group of the free Schr\"odinger
equation'' \cite{Nie}.
The Schr\"odinger group is thus isomorphic to
$(SO(3)\times SL(2,{\bf R}))\,
{\ooalign{\hfil\raise.07ex\hbox{s}\hfil\crcr\mathhexbox20D}} 
\,({\bf R}^3\times{\bf R}^3)$,
i.e. to the multiplicative group of those $5\times 5$ matrices
\begin{equation}
\pmatrix{
A & {\bf b} & {\bf c} \cr
{\bf 0} & d & e \cr
{\bf 0} & f & g \cr
}
\label{4.9}
\end{equation}
with entries as above.

iv) The $14$-dimensional group of conformal automorphisms of the flat Bargmann
structure, the ``chronoprojective group'' \cite{PBD} that preserves
the {\em directions} of $g$ and $\xi$ separately (see (\ref{4.3},\ref{4.4})),
can be thought of as a preferred Lie subgroup of $O(5,2)$; its commutator
subgroup turns out to be the extended Schr\"odinger group (\ref{4.7}).

v) Finally, the $3$-parameter subgroup
$(A = {\bf 1},{\bf b} = {\bf c} = {\bf 0})$ of the Schr\"odinger group
(\ref{4.9}) is the group of {\em non-relativistic conformal transformations}
(\ref{1.11}--\ref{1.13}) isomorphic to $SL(2,{\bf R})$ interpreted as the group of
projective transformations of the time axis. The dilatations and expansions
introduced in Sec.~1 form the triangular Borel subgroup $(A = {\bf 1},{\bf b} =
{\bf c} = {\bf 0},e = 0)$.

\noindent
{\bf Remark 1.} Comparing with (\ref{3.9}) we conclude that, for each
value of the parameters $a,b,c$, the spacetime projection of the
Vinti-Lynden-Bell transformation (\ref{3.7}) belongs to the $SL(2,{\bf R})$ subgroup.
This can also be understood by observing that for an element $\cal C$ of the
extended Schr\"odinger group, one has
${\cal C}^* g_U = \Omega^2 g_0 - 2\,{\cal C}^* (U dt \otimes dt)$,
see (\ref{2.1}) and (\ref{4.1}). Thus, Eq. (\ref{3.3}) is satisfied as soon as
\begin{equation}
{\cal C}^*(U dt \otimes dt) = \Omega^2 U_0\,dt \otimes dt.
\label{4.10}
\end{equation}
A closer inspection shows that this corresponds to the calculation of Sec.~3.

\noindent
{\bf Remark 2.} In the same spirit, one can ask which potentials $U$ are
conformally related to the free case, namely
\begin{equation}
{\cal A}^* g_U  = g_0,
\qquad\hbox{and}\qquad
{\cal A}_* \xi = \xi.
\label{4.11}
\end{equation}
A short calculation yields \cite{Duv2}
\begin{equation}
U({\bf q},t) =
\mbox{$\textstyle{1\over 2}$}
u(t) {\bf q}^2 + \langle {\bf v}(t),{\bf q} \rangle + w(t),
\label{4.12}
\end{equation}
where $u,{\bf v},w$ are arbitrary functions of time. Physically, we can have a
time-dependent spherically symmetric {\em harmonic oscillator} plus a
{\em homogenous force}.
This explains why the classical symmetries of these systems are related to those
of the free particle.
This fact has been exploited in the quantum mechanical framework \cite{BDP}
to conformally relate the general solutions of the free Schr\"odinger
equation to those of the Schr\"odinger equation in the presence of a potential
of the form (\ref{4.12}), e.g. a (time-dependent) harmonic oscillator.

\noindent
{\bf Remark 3.} One could also ask which central potentials $U$ are invariant
with respect to non-relativistic conformal transformations given by Eqs
(\ref{1.11}--\ref{1.13}). Due to (\ref{4.10}), this requires
\begin{equation}
U({\bf q},t) = {\mathop{\rm const}\nolimits\over{\bf q}^2}.
\label{4.13}
\end{equation}
A generalization of this result to the case of $N$ bodies has been obtained
in Ref.~\cite{BP}.

Finally, adding a Dirac magnetic monopole would only change the symplectic
structure introduced in Sec.~2 by a term proportional to the area $2$-form of
$S^2$, which is manifestly invariant with respect to our non-relativistic
conformal transformations.

The most general conformally invariant system is thus an {\em inverse square
potential\/} \cite{AFF} plus a {\em Dirac monopole} \cite{Jac1,Duv1,Hor}.


\section{The $N$-body problem}

In non-relativistic physics, it is consistent to confine attention to a
finite number, $N$, of bodies moving in Euclidean space ${\bf R}^3$.
An equivalent description is to give a curve in the configuration space
${\bf R}^{3N}$. To obtain a spacetime description, we may then add one extra
absolute time variable to obtain a Newtonian spacetime of dimension $3N + 1$.
The motion of the bodies corresponds to a worldline in this $N$-body
space\-time.

\subsection{The $N$-body Bargmann structure}

The metric of this $(3N + 2)$-dimensional Bargmann structure is
\begin{equation}
g_V \equiv \sum_{j=1}^N{
{m_j \over m} \langle d{\bf q}_j \otimes d{\bf q}_j \rangle
}
+ dt \otimes ds + ds \otimes dt
- {2\over m} V({\bf q}_1,\ldots,{\bf q}_N,t)\,dt \otimes dt,
\label{5.1}
\end{equation}
where
$m_1,\ldots,m_N$ are the masses of the bodies and $m = m_1 + \cdots + m_N$ is 
the total mass of the system.
As before
\begin{equation}
\xi \equiv {\partial\over \partial s}
\label{5.2}
\end{equation}
is the $({\bf R},+)$-generator that defines the principal null, covariant-constant
fibration.

For the planetary $N$-body problem of celestial mechanics, we take
\begin{equation}
V({\bf q}_1,\ldots,{\bf q}_N,t) =
- \sum_{j<k}{G\,{m_j m_k \over \| {\bf q}_j - {\bf q}_k \|}}
\label{5.3}
\end{equation}
and thus define the metric in
$Q = ({\bf R}^{3N} \backslash \Delta) \times {\bf R} \times {\bf R}$
where $\Delta$ is the collision subset.

Note that the potential $V$ in (\ref{5.3}) consistently leads to a Ricci-flat
metric $g_V$ given by (\ref{5.1}), i.e. a  solution of the vacuum field equations~(\ref{3.1}).

The non zero Christoffel symbols are
$\Gamma_{tt}^{A_j} = (1/m_j)\partial_{A_j}V,\,
\Gamma_{{A_j}t}^{s} = -(1/m)\partial_{A_j}V$
and
$\Gamma_{tt}^{s} = -(1/m)\partial_{t}V$
where we have put
${\bf q}^{A_j} \equiv {\bf q}^A_j$ with $A = 1,2,3\ \&\ j = 1,\ldots,N$.
The equations of the null geodesics are readily interpreted as Newton's
equations of motion, viz.
\begin{equation}
{d^{2}{\bf q}^{A_j} \over dt^{2}} = - {1 \over m_j}\partial_{A_j}V,
\label{5.4}
\end{equation}
together with a supplementary equation for the action
\begin{equation}
{d^2 s \over dt^2} = {1 \over m} \left(2\,{dV \over dt}
- {\partial V \over \partial t}\right).
\label{5.5}
\end{equation}
Note that the metric $g_V$ is indeed conformally defined by its null geodesics,
i.e. the solutions of the $N$-body equations of motions. In other terms, the
``shape'' of the extended spacetime is defined, up to a factor, by the
motions of matter in the universe.

Regarding the time-variation of the ``constant'' of gravitation, our
previous arguments for the external Newtonian field apply here just as well. We
can therefore claim that the only gravitational ``constant'' $G(t)$ in
(\ref{5.3}) that can be $SO(3)$-equivariantly associated with $G_0$ is again
given by Vinti's formula (\ref{3.6}).

\goodbreak

We remark {\em en passant} that, had we been considering the analogous problem
in electrostatics with a Coulomb's potential
\[
V({\bf q}_1,\ldots,{\bf q}_N,t) =
\sum_{j<k}{e_j e_k \over \|{\bf q}_j - {\bf q}_k\|}
\]
for a system of charges $e_1,\ldots,e_N$, we would have found exactly the same
possible time-dependence for for the fine structure constant
\begin{equation}
e^2 \propto {1 \over t}.
\label{5.7}
\end{equation}

\subsection{Homographic solutions}

The only known general class of non planar exact solutions of the $N$-body
problem are the so-called homographic solutions. Their existence is related to
the conformal structure of the Bargmann manifold. These particular solutions
have the form
\begin{equation}
{\bf q}_j(t) = \Omega(t)\,{\bf q}_j^0.
\label{5.8}
\end{equation}
Substituting this Ansatz into the equations of motion (\ref{5.4}), where $V$ is
given by (\ref{5.3}) with $G = \mathop{\rm const}\nolimits$, yields
\begin{equation}
{\bf q}_j^0 = {1\over \lambda}
\sum_{k \ne j} Gm_k {{\bf q}_j^0 - {\bf q}_k^0 \over \|{\bf q}_j^0 - {\bf q}_k^0\|^3}
\qquad\hbox{with}\qquad
\lambda = - \Omega^2\ddot\Omega.
\label{5.9}
\end{equation}
The solutions ${\bf q}^0 = ({\bf q}_1^0,\ldots,{\bf q}_N^0)$ of these  equations (with
$\lambda = \mathop{\rm const}\nolimits > 0$) are called {\em central configurations} and provide  by
(\ref{5.8}) some exact solutions of the $N$-body problem which, up to now, have
not been completely classified (see Ref.~\cite{CL} for an account of recent
progress in the classification of the non-planar central configurations with
equal masses).

To gain some further insight, let us first observe that if we set
$dt^0 = dt/\Omega(t)^2$ and, of course, ${\bf q}_j^0 = {\bf q}_j/\Omega(t)$, then
$$
\left({d{\bf q}_j \over dt}\right)^2\,dt
=
\left(
\left({d{\bf q}_j^0 \over dt^0}\right)^2 +
\lambda\Omega\,{{\bf q}_j^0}^2
\right) dt^0
+ d\left(\Omega\dot\Omega\,{{\bf q}_j^0}^2 \right)
$$
and
$$
V({\bf q})\,dt = \Omega V({\bf q}^0)\,dt^0.
$$
Thus, up to a total derivative, we get the Lagrangian of $N$ particles
interacting with a combined repulsive oscillator and Newtonian gravitational
field (with time-dependent coefficients). But this latter system admits a static
equilibrium configuration, namely when the gravitational attraction is cancelled
by the linear repulsion.

Again, this can be rephrased geometrically in terms of conformal transformations
of the $N$-body Bargmann manifold $(Q,g_V^0,\xi^0)$ with
$Q = ({\bf R}^{3N}\backslash\Delta) \times{\bf R} \times{\bf R}$ and 
$
g_V^0 = \sum_{j=1}^N{(m_j/m) \langle d{\bf q}_j^0 \otimes d{\bf q}_j^0 \rangle}
+ dt^0 \otimes ds^0 + ds^0 \otimes dt^0
- (2/m) V({\bf q}^0)\,dt^0 \otimes dt^0
$
and
$
\xi^0 = \partial/\partial s^0
$
with $V$ given by (\ref{5.3}).
A simple calculation, akin to the previous remark, shows that the mapping
${\cal A} : ({\bf q}^0,t^0,s^0) \to ({\bf q},t,s)$ of $Q$, whose inverse is given by
\begin{equation}
\left\{
\begin{array}{l}
{\bf q}^0 = {\displaystyle {{\bf q}\over\Omega(t)}}\\[8pt]
t^0 = {\displaystyle \int\!{dt\over\Omega(t)^2}}\\[6pt]
s^0 = {\displaystyle
s + {\dot\Omega\over 2m\Omega}\sum_{j=1}^N{m_j {\bf q}_j^2},
}
\end{array}
\right.
\label{5.10}
\end{equation}
transforms the original Bargmann structure according to
\begin{equation}
g_V \equiv ({\cal A}^{-1})^*g_V^0 = \Omega^2\,g_{V_{\rm eff}}^0
\qquad\hbox{and}\qquad
\xi \equiv {\cal A}_*\xi^0 = \xi^0
\label{5.11}
\end{equation}
where
\begin{equation}
V_{\rm eff}({\bf q}^0,t^0) = \Omega \left(V({\bf q}^0)
+ \mbox{$\textstyle{1\over 2}$}
\Omega^2\ddot\Omega \sum_{j=1}^N{m_j {{\bf q}_j^0}^2}\right).
\label{5.12}
\end{equation}

The critical points ${\bf q}^0$ of $V_{\rm eff}$ are the static equilibria (or the
central configurations), i.e. the solutions of
$\mathop{\bf grad}\nolimits_jV_{\rm eff}({\bf q}^0,t^0) = {\bf 0}$
(implying $\lambda = - \Omega^2\ddot\Omega = \mathop{\rm const}\nolimits$).
These actually define some specific null geodesics of $g_{V_{\rm eff}}^0$ which,
according to Eq. (\ref{5.11}) happens to be conformally related to $g_V$. We
again note (\ref{5.11}) that the total mass is preserved by the transformation
$\cal A$. Central configurations are indeed associated with null geodesics of
$g_V$, hence to some particular solutions $(t \to ({\bf q}_1(t),\ldots,{\bf q}_N(t)))$
of the original set (\ref{5.4}) of Newton's equations.

\noindent
{\bf Remark.} A similar explanation can be given to the observation of
Forg\'acs and Zakrzewski \cite{FZ} who found that the action 
$\int\!{f(t)\dot y^2(t)\,dt}$
can be brought into that of a free particle by the change of variable
$t \to \int\!{dt/f(t)}$.

\subsection{The (cosmic) virial theorem}

We conclude this section with a remark about scaling and the {\em virial
theorem}. 

Consider the Newtonian $N$-body problem described by the metric $g_V$ in
(\ref{5.1}) with $V$ as in (\ref{5.3}) and $G = \mathop{\rm const}\nolimits$. This Bargmann structure
admits a $5$-dimensional Lie algebra of fibre preserving conformal Killing
vectors. This algebra consists of $4$ isometries (rotations and time
translations) and  of the homo\-thetic-Killing vector field 
\begin{equation}
X = \sum_{j=1}^N{{\bf q}_j\cdot{\partial\over\partial {\bf q}_j}} 
+ {3\over 2} t\,{\partial\over\partial t} 
+ {1\over 2} s\,{\partial\over\partial s}.
\label{5.13}
\end{equation}

\goodbreak

This latter generates the {\em homothety group} $(\Lambda > 0)$ :

\begin{equation}
\left\{
\begin{array}{l}
{\bf q}_j \to {\bf q}_j^* = \Lambda {\bf q}_j\\[6pt]
t \to t^* = \Lambda^{3/2} t\\[6pt]
s \to s^* = \Lambda^{1/2} s,
\end{array}
\right.
\label{5.14}
\end{equation}
with $j = 1,\ldots,N$, under which $g_V \to \Lambda^2 g_V$ and
$\xi \to \Lambda^{-1/2}\xi$.
Thus  
\begin{equation}
L_X\,g_V = 2\,g_V
\qquad\hbox{and}\qquad
L_X\,\xi = - \mbox{$\textstyle{1\over 2}$} \,\xi.
\label{5.15}
\end{equation}

Such homotheties lift to the cotangent bundle $(T^*Q,\vartheta)$ as canonical
symplectic similitudes $(\vartheta^* = \Lambda^{1/2} \vartheta)$, namely as
\begin{equation}
\left\{
\begin{array}{l}
p_{A_j}^* = \Lambda^{-1/2} p_{A_j} \\[6pt]
p_t^* = \Lambda^{-1} p_t \\[6pt]
p_s^* = p_s,
\end{array}
\right.
\label{5.16}
\end{equation}
where $A_j = 1,2,3\ \&\ j=1,\ldots,N$.
These homotheties (\ref{5.16}) preserve therefore the $(6N+2)$-dimensional
submanifold $C$ defined by the constraints (\ref{2.4},\ref{2.5}) and the null
foliation $\ker(d\vartheta_C)$. Hence, they permute the classical motions. This
yields a {\em generalized Kepler's third law} : if $(t \to {\bf q}(t))$ is a
solution of Newton's equations for $N$ bodies then so is $(t \to
\Lambda{\bf q}(\Lambda^{-3/2}t))$.

As a by-product, we get the {\em virial theorem} used by astrophysicists to
estimate the mass of clusters of galaxies \cite{Pee}.
We have seen in fact that, if $\,\overline{\!X}$ denotes the canonical lift
of the vector field $X$ to $T^*Q$, then
\begin{equation}
L_{\overline{\!X}}\,\vartheta = 
\mbox{$\textstyle{1\over 2}$}
\vartheta,
\label{5.17}
\end{equation}
and thus if $Y \in \ker (d\vartheta_C)$ is a generator of the equations
of motion (\ref{2.6}), we have
$Y(\vartheta_C(\overline{\!X})) = {1 \over 2}\vartheta_C(Y)$.
Using (\ref{2.6}), $- p_t = E$ (energy) and $p^t = m$ (total mass), this can be
rewritten as
$
Y\left(\sum{p_{A_j} q^{A_j}}\right) - {3\over 2} Em\alpha = 
-{1\over 2}m\alpha{}p^s.
$
Introducing the kinetic energy $T = \sum{p_{A_j} p^{A_j}}/(2m_j)$ we get,
since $p^s=-E+2V$,
\begin{equation}
2T+V = {1 \over m\alpha} Y \left(\sum{p_{A_j} q^{A_j}}\right).
\label{5.18}
\end{equation}
If we assume that the system is in equilibrium then the average of the
RHS of (\ref{5.18}) vanishes and we are left with
\[
2T =-V \quad\hbox{(time average)}.
\]

All these results can be extended to include the case of a time-dependent
gravitational ``constant'' $G(t)$. As we previously mentioned in Sec.~3,
this case can also be reinterpreted as giving the equations of motion for
particles in an expanding universe (Eq. (\ref{1.4}) if we use the interpretation
(\ref{1.5})). The virial theorem (\ref{5.18}) then becomes the ``cosmic
virial theorem'' \cite{Pee}. If $G$ is not constant with time, $\partial/\partial
t$ will no longer be a Killing vector field of the metric $g_V$ given in
(\ref{5.1}). In fact, \begin{equation}
\left(L_{\partial/\partial t}\,g_V\right)_{ab} = 
- {2\over m} {\dot G(t)\over G(t)} \,V \xi_a\xi_b.
\label{5.20}
\end{equation}
Using this equation, it is easy to obtain the ``cosmic energy
equation'' \cite{Pee}.


\section{Hidden symmetries and Killing tensors}

We now briefly describe the appearance of ``hidden'' symmetries. 

In our formalism, a {\em manifest symmetry} belongs to the group of conformal
auto\-morphisms of the bundle $(Q,g,\xi)$, whose infinitesimal generators are the
{\em conformal-Killing vector fields} $\kappa$ that commute with the 
$({\bf R},+)$-generator, namely $(a,b=1,\ldots,5)$ :
\begin{equation} 
\nabla_{(a} \kappa_{b)} = \lambda\,g_{ab} \qquad\hbox{and}\qquad
\lbrack \xi,\kappa \rbrack = 0,
\label{6.1}
\end{equation}
for some function $\lambda$ on $Q$. (Round brackets denote symmetrization.)

The functions $H_\kappa = p_a \kappa^a$, linear in momentum on $T^*Q$, are
constants of the motion of our test particle.

For example, in the case of the $1$-body problem with $G = G_0$, rotations,
time-translations and vertical translations act by isometries and the
associated conserved quantities are respectively the {\em angular momentum} 
${\bf L}$, the {\em energy} $E = - H_{\partial/\partial t}$ and the {\em mass}
$m = H_\xi$.

Now, the so called ``accidental'' or ``hidden'' symmetries
associated with the Lagrange-Laplace-Runge-Lenz vector can also be discussed
in this new setup. Observe that the {\em quadratic quantities}
\begin{equation}
H_\kappa = 
\mbox{$\textstyle{1\over 2}$}
\,p_a p_b \kappa^{ab}
\label{6.2}
\end{equation}
are conserved along null geodesics of $(Q,g)$ whenever
\begin{equation}
\nabla_{(a} \kappa_{bc)} = \lambda_{(a} g_{bc)}
\label{6.3}
\end{equation}
for a symmetric and tracefree tensor $\kappa$ (see e.g. Refs~\cite{BF,GRF}).
These objects are called {\em con\-formal-Killing $2$-tensors}. Since we want to
preserve the principal bundle structure on the extended spacetime $(Q,\xi)$, we
will only deal with con\-formal-Killing tensors $\kappa$ such that
\begin{equation}
L_\xi\,\kappa = 0,
\label{6.4}
\end{equation}
i.e. projectable ones.  


In the Kepler case, a lengthy calculation is needed to prove that the following
expression is indeed a solution of the Killing equations (\ref{6.3}) and
(\ref{6.4}) :
\begin{equation}
\kappa^{ab} = \eta^{ab} -
\mbox{$\textstyle{1\over 5}$}
\hat\eta g^{ab}
\qquad\hbox{with}\qquad
\hat\eta = \eta^{ab}g_{ab},
\label{6.5}
\end{equation}
where the nonvanishing contravariant components of $\eta$ are given by
\begin{equation}
\eta^{AB} = \omega^A q^B + \omega^B q^A - \hat\eta\,\delta^{AB}
\label{6.6}
\end{equation}
with $A,B,C = 1,2,3$ and
\begin{equation}
\eta^{45} = \eta^{54} = \hat\eta \ (= \omega_C q^C)
\label{6.7}
\end{equation}
for some $\mbox{\boldmath$\omega$} \in {\bf R}^3$.

It is finally easy to check, with the help of Eq. (\ref{6.2}), that 
$H_\kappa = \langle\mbox{\boldmath$\omega$},{\bf A}\rangle$ where
\begin{equation}
{\bf A} = {\bf L} \times {\bf p} + m^2 m_0 G_0 {{\bf q} \over r}
\label{6.8}
\end{equation}
is the {\em Lagrange-Laplace-Runge-Lenz vector}. See Ref.~\cite{Cra} for an
alternative treatment in the $4$-dimensional setting.


\section{Relation to the work of Brinkmann \& Kaluza-Klein theory}

We now wish to point out the relation of our results to the work of
Brinkmann \cite{Bri} who discussed the circumstances under which two metrics $g$
and $g^*$ related by a conformal rescaling,
\[
g^* = \Omega^2 g,
\]
might both be Einstein, i.e. both satisfy
\[
R_{ab} = \Lambda\,g_{ab}.
\]
He distinguished three cases,
\[
\begin{array}{ll}
A) \quad \Omega = \mathop{\rm const}\nolimits, \\[6pt]
B) \quad g^{ab}\,\partial_a \Omega\,\partial_b \Omega \ne 0, \\[6pt]
C) \quad g^{ab}\,\partial_a \Omega\,\partial_b\Omega = 0,
\qquad \partial_a \Omega \ne 0.
\end{array}
\]
It is case (C), called ``improper conformal rescalings'', which is relevant
for us since our coordinate $t$ and hence any function of it satisfies condition
(C). Brinkmann included the possibility that $g^*$ was the pull-back of $g$
under a diffeomorphism so his results are directly applicable. He showed that in
case (C), $g$ and $g^*$ must admit a covariantly constant null vector field.
Thus, in particular, they must be Ricci flat. In other words he established that
$g$ and $g^*$ must admit what we have referred to as a {\em Bargmann structure}.
This implies a uniqueness property of the Vinti-Lynden-Bell transformations.

Brinkmann then went on to determine explicitly all Einstein-Bargmann
structures in dimensions $4$ and $5$ and to give a set of necessary and
sufficient conditions in all higher dimensions.

\goodbreak

\subsection{Case $D = \mbox{4}$}

In $4$ dimensions the most general Einstein-Bargmann structure is expressed as
\begin{equation}
g \equiv dq^1 \otimes dq^1 + dq^2 \otimes dq^2
+ dt \otimes ds + ds \otimes dt
- 2 U(q^1,q^2,t)\,dt \otimes dt
\label{7.4}
\end{equation}
with 
\begin{equation}
\xi \equiv {\partial\over\partial s}
\label{7.5}
\end{equation}
and
\begin{equation}
\Delta\,U \equiv \left(\partial_1^2 + \partial_2^2 \right)\,U = 0.
\label{7.6}
\end{equation}
\goodbreak
Metrics of this form are referred to as {\em plane-fronted gravitational waves
with parallel rays} (or pp-waves) in the general relativity literature. The
special cases when $U$ is quadratic in the ${\bf q}$'s are called {\em exact plane
gravitational waves}. They admit a $5$-parameter group of isometries acting on
the null $3$-surfaces  $t = \mathop{\rm const}\nolimits$.

\subsection{Case $D = \mbox{5}$}

In $5$ dimensions, the most general Einstein-Bargmann structure is, ac\-cording
to Brink\-mann, of the form
\begin{eqnarray}
g \equiv & &\langle d{\bf q} \otimes d{\bf q} \rangle
+ dt \otimes (ds + {\bf A}.d{\bf q})
+ (ds + {\bf A}.d{\bf q}) \otimes dt \nonumber\\ 
& & + \left(-2U +
\mbox{$\textstyle{1\over 2}$}
V^2 \right)\,dt\otimes dt,
\label{7.7}
\end{eqnarray}
with $\xi$ just as before; ${\bf A},U,V$ being functions of ${\bf q}\ \&\ t$
such that
\begin{equation}
\mathop{\bf curl}\nolimits\,{\bf A} = \mathop{\bf grad}\nolimits\,V
\qquad\hbox{and}\qquad
\Delta\,U = 0.
\label{7.8}
\end{equation}

i) If $V = 0$, we obtain the obvious generalization to $5$ spacetime dimensions
of plane-fronted waves.

ii) If $U$ is taken to be quadratic in the ${\bf q}$'s, the metric admits in general
a $7$-parameter group of isometries acting on the surfaces $t = \mathop{\rm const}\nolimits$,
hence preserving the fibration.

iii) In the flat case, $U = V = 0,{\bf A} = {\bf 0}$, the group of isometries
which preserve the slices $t = \mathop{\rm const}\nolimits$ is larger; it may be viewed as the
commutator subgroup of the Bargmann group called the {\em Carroll group},
isomorphic to the $10$-dimensional Lie group of all $5\times 5$ matrices
\begin{equation}
\pmatrix{
A &{\bf 0} &{\bf c} \cr
-{}^t{\bf b}A &1 &h \cr
{\bf 0} &0 &1 \cr
}
\label{7.9}
\end{equation}
(with $A\in SO(3);{\bf b},{\bf c}\in {\bf R}^3;h\in {\bf R}$), acting on the 
``position-action'' affine plane spanned by
$$
\pmatrix{
{\bf q}\cr
s\cr
1\cr
}
$$ 
as deduced from (\ref{4.7}) where we have set $t = 0$ with
$d = g = 1$ and $e = f = 0$.
It is worth remembering that the Carroll group has been orginally introduced by
L\'evy-Leblond as the contraction $c \rightarrow 0$ of the Poincar\'e
group \cite{LL2}.

iv) If we set $V = 0$ and $U$ given by Eq. (\ref{3.2}), we obtain the Bargmann
structure associated with a single Newtonian point particle.

v) If $V \ne 0$ we obtain something more, namely a generalization of
Newtonian non-relativistic theory which, although not envisaged in the usual
elementary treatments, occurs in the formulation due to Cartan \cite{Car,Kun}.
In fact, Cartan's version of the theory is not merely a reformulation but an
extension since it allows a new phenomenon, the possibility of {\em magnetic
mass} \cite{DK0}.
In General Relativity, this possibility was recognized first in the Taub-NUT
solutions.
By taking a suitable limit as $c \rightarrow \infty$, Koppel \cite{Kop} found a
new solution of Newton-Cartan field equations on spacetime. Here, we interpret
it as the ``non-relativistic Taub-NUT solution'' of the vacuum field
equations (\ref{3.1}) corresponding to the Bargmann structure
$(({\bf R}^3\backslash\{0\} \times {\bf R}) \times {\bf R},g,\xi)$ where $g$ is given by
(\ref{7.7}) and
\begin{equation}
U({\bf q}) = - G {m_0 \over \|{\bf q}\|},\qquad
{\bf A}.d{\bf q} = - \ell\,\cos \theta\,d\phi,\qquad
V({\bf q}) = - {\ell \over \|{\bf q}\|}.
\label{7.10}
\end{equation}
The vector field $\xi$ is still given by (\ref{7.5}) and $\ell$ is a new constant
homogeneous to an [action]/[mass].
To interpret $V$, we note that the equations of motion of a test particle
read in this case
\begin{equation}
{d^2 {\bf q} \over dt^2} = - \mathop{\bf grad}\nolimits\left(U -
\mbox{$\textstyle{1\over 4}$}
V^2 \right)
+ {d{\bf q} \over dt} \times \mathop{\bf grad}\nolimits\,V,
\label{7.11}
\end{equation}
thus $\ell$ corresponds to a Newtonian magnetic mass monopole.
The solution (\ref{7.10}) is clearly singular at the origin ${\bf q} = {\bf 0}$
for all non zero values of the two parameters $m_0$ and $\ell$. 

It is worth mentioning that the gravitational ``constant'' $G$ could 
actually depend on time. If $G$ varies inversely as time while $\ell$ remains a
constant, the Vinti-Lynden-Bell transformation (\ref{3.7}) still brings the
system into a time-independent form. This is so because both the monopole term
coming from $\mathop{\bf curl}\nolimits\,{\bf A}$ and the $\ell^2/r^2$ term 
coming from $V^2$ are symmetric with respect to non-relativistic conformal 
transformations. The metric (\ref{7.7},\ref{7.10}) falls short of having an 
extra ``hidden'' symmetry :
if one had $U - V^2/2$ rather than $U - V^2/4$ in (\ref{7.11}), the metric would
admit a conformal Killing tensor \cite{GRF} yielding a conserved
Lagrange-Laplace-Runge-Lenz vector.

\goodbreak

The $1$-form
\begin{equation}
\varpi \equiv ds - \ell\cos\theta\,d\phi
\label{7.12}
\end{equation}
appearing in the metric (\ref{7.7}) turns out to be directly related to the
canonical connection $\alpha$ living on the Hopf circle bundle $S^3 \to S^2$,
viz.
\[
\alpha \equiv {\varpi\over 2\ell},
\]
provided $s$ is taken to be periodic with period $4\pi\ell$.
The associated parameters $(\theta,\phi,\psi)$, with 
$\psi\equiv s/(2\ell) \pmod{2\pi}$, are the Euler angles on $S^3$ whilst our
Taub-NUT like Bargmann structure $(Q,g,\xi)$ is now globally defined by the
$SO(2)$-bundle
$
Q\cong (S^3 \times {\bf R}^+) \times {\bf R}
\to
B\cong (S^2 \times {\bf R}^+) \times {\bf R}
$
endowed with the metric
\begin{equation}
g = dr\otimes dr + r^2 g_{S^2}
+ 2\ell\,(\alpha\otimes dt + dt\otimes \alpha)
+ \left(2G\,{m_0 \over r} + {\ell^2 \over 2r^2}\right)\,dt\otimes dt
\label{7.14}
\end{equation}
and the circle-generator
\begin{equation}
\xi = {1\over 2\ell}{\partial\over\partial\psi}.
\label{7.15}
\end{equation}

However, by contrast with the relativistic Taub-NUT solution in which the
relativistic time must be periodic, the Newtonian time variable need not be
periodic in our case. The periodicity has an interesting quantum-mechanical
consequence which we will discuss in Sec.~8.

\noindent
{\bf Remarks.}
When $U$, $V$ and $\bf A$ are independent of $t$, the metric
(\ref{7.7},\ref{7.8})  admits an additional Killing vector field,
$\partial/\partial t$, which, if  $-2U + {1 \over 2} V^2 > 0$, will be
spacelike. This allows us to give the $5$-metric a conventional Kaluza-Klein
interpretation. If $V = 0$, and
\begin{equation}
- 2 U({\bf q}) = 1 + \sum_{j=1}^{N}{X_j \over \|{\bf q} - {\bf q}_j\|}, 
\label{7.16}
\end{equation}
one obtains a metric representing $N$ point-particles in equilibrium \cite{Gib}.
The masses, scalar charges and electric charges are in the ratio $1 : \sqrt{3} :
2$ which implies that the gravitational and scalar attractions are exactly
cancelled by the electrostatic repulsion. These metrics are in a certain sense
dual to the multi-Taub-NUT metrics which have an interpretation as Kaluza-Klein
monopoles \cite{GR,GW}. The case $N = 1$ may be obtained from the $4$-dimensional
Schwarzschild metric by boosting it in the $5$th direction up to the speed of
light \cite{Gib}.

In the case where $V \ne 0$, we obtain metrics whose Kaluza-Klein
inter\-pretation is that of particles with both electric and magnetic
charges, i.e. {\em Kaluza-Klein dyons}.

\subsection{Case $D \geq \mbox{5}$}

Let us complete our review of Einstein-Bargmann structures by giving
Brinkmann's result for spacetime dimensions $D$ exceeding $5$.

The metric takes the form $(A,B,\ldots = 1,2,\ldots,D-2)$ :
\begin{equation}
g \equiv \widehat g_{AB}({\bf q},t)\,dq^A \otimes dq^B
+ dt \otimes \varpi + \varpi \otimes dt 
+ H({\bf q},t)\,dt \otimes dt
\label{7.17}
\end{equation}
where
$$
\varpi \equiv ds + A_K({\bf q},t)\,dq^K
$$
and
\begin{equation}
\left\{
\begin{array}{ll}
R_{AB} &= \widehat R_{AB} = 0 \\[6pt]

R_{At} &= - \frac{1}{2}
\left[\partial_A \left(\widehat g^{KL}\partial_t \widehat g_{KL}\right)
+ \widehat\nabla^K F_{KA}
\right] = 0 \\[6pt]

R_{tt} &= - \frac{1}{2}
\left[\widehat\Delta\,H
+ \partial_t \left(\widehat g^{KL}\partial_t \widehat g_{KL}\right)
+ \frac{1}{2} F^{KL} F_{LK}
- 2 \widehat\nabla^K \partial_t A_K
\right] = 0.
\end{array}
\right.
\label{7.18}
\end{equation}
Here $F$ is defined to be
\begin{equation}
F_{KL} \equiv \partial_K A_L - \partial_L A_K - \partial_t \widehat g_{KL}.
\label{7.19}
\end{equation}

We display, for completeness, the associated non-trivial Christoffel symbols :
\begin{equation}
\left\{
\begin{array}{ll}
\Gamma^A_{BC} &= \widehat\Gamma^A_{BC} \\[4pt]

\Gamma^A_{Bt} &= -\frac{1}{2}\,\widehat g^{AK}F_{KB} \\[4pt]

\Gamma^A_{tt} &= \widehat g^{AK}\left(\partial_t A_K 
- \frac{1}{2}\partial_K H\right) \\[4pt]

\Gamma^s_{BC} &= 
\partial_{(B} A_{C)} - A_K \widehat\Gamma^K_{BC} 
- \frac{1}{2}\partial_t \widehat g_{BC} \\[4pt]

\Gamma^s_{Bt} &= \frac{1}{2}\,\widehat g^{KL} F_{KB} A_L
+ \frac{1}{2}\partial_B H \\[4pt]

\Gamma^s_{tt} &= \frac{1}{2}\,
\widehat g^{KL} A_K \left(\partial_L H - 2\partial_t A_L\right) 
+ \frac{1}{2}\partial_t H.
\end{array}
\right.
\label{7.20}
\end{equation}

The metric (\ref{7.17}) is likely to have a number of applications to
Kaluza-Klein supergravity and superstring theory. We could, for instance, take
$\widehat g$ to be the metric on a Calabi-Yau space. We defer discussion of
these possibilities to another time.

\goodbreak

We would like to finish this section by giving a Taub-NUT like exact solution for
the Newtonian $N$-body field equations. It actually generalizes the preceding
solution (\ref{7.14},\ref{7.15}) as well as the classical ``inverse square
law'' (\ref{5.1}--\ref{5.3}) to a situation where the $N$ massive bodies are
allowed to carry an additional ``magnetic mass'', viz. {\em gravitational
dyons}. The corresponding Ricci-flat metric of the Bargmann manifold we are
dealing with in dimension $D = 3N + 2$, is of the general form (\ref{7.17}),
namely
\begin{equation}
g \equiv \sum_{j=1}^N{
{m_j \over m} \langle d{\bf q}_j \otimes d{\bf q}_j \rangle
}
+ dt \otimes \varpi + \varpi \otimes dt
+ H({\bf q})\,dt \otimes dt,
\label{7.21}
\end{equation}
with
\begin{equation}
\varpi \equiv ds + \sum_{j<k}
{\epsilon_{jk}\,
{\langle{\bf u}_{jk}\times{\bf q}_{jk},d{\bf q}_{jk}\rangle
\over
r_{jk}^2 + r_{jk}\,\langle{\bf u}_{jk},{\bf q}_{jk}\rangle}
}
\label{7.22}
\end{equation}
and
\begin{equation}
H({\bf q}) = {2G\over m}
\sum_{j<k}{{m_j m_k \over r_{jk}}}
+ \sum_{j=1}^N{{m\over 2m_j}
\left(\sum_{k\ne j}{{\epsilon_{jk} \over r_{jk}}}\right)^2}.
\label{7.23}
\end{equation}
The notations are as follows :
${\bf q}_{jk} \equiv {\bf q}_j - {\bf q}_k,\,
r_{jk} \equiv \|{\bf q}_{jk}\|$
with $j,k = 1,\ldots,N$, while the unit vector ${\bf u}_{jk} = \mathop{\rm
const}\nolimits$ defines  an otherwise arbitrary direction ---the direction of
the Dirac string--- entering the {\em local} expression of the $1$-form
(\ref{7.22}). Finally, the masses $m_j$ and magnetic masses $\ell_j$ are
skew-symmetrically encoded into the coefficients \begin{equation}
\epsilon_{jk} \equiv {\ell_j m_k - \ell_k m_j \over m}
\label{7.24}
\end{equation}
where $m \equiv \sum_{j=1}^N{m_j}$.
Again, the global differential structure of this Bargmann manifold can be
worked out in a similar (although more involved) way as before.


\section{Quantization}

We are now ready to derive the quantum version of the preceding classical
results. We follow the general rules given by De Witt \cite{DeW} of quantization
in a curved manifold. In dealing with topologically non trivial solutions of
Newtonian gravity, we will resort to prequantization \cite{JMS} in order to
establish a mass quantization formula analogous to the Dirac quantization
formula for the electric charge.

\subsection{The covariant Schr\"odinger equation}

The kinetic energy $g^{ab}{p_a}{p_b}$ (\ref{2.4}) is quantized as 
$
-\hbar^2 (\Delta_g - {1 \over 6} R_g),
$
where $\Delta_g$ is the Laplace operator on $(Q,g)$ and $R_g$ is the scalar
curvature ---we have identically $R_g = 0$ as a consequence of Newton's field
equations (\ref{3.1}). This result can also be obtained by
geometric quantization \cite{Sni} using the vertical polarization of
$(T^*Q,\omega)$. It is consistent with the quantization rule of Killing
tensors given in (\ref{8.10}).

The mass Hamiltonian $p_a\xi^a$ (\ref{2.5}) is linear in momentum, its
quantization is therefore canonical. By quantizing the constraints
(\ref{2.4},\ref{2.5}) according to Dirac's prescription, we obtain the
following set of PDE's on the Bargmann manifold $(Q,g,\xi)$ :
\begin{equation}
\Delta_g \psi = 0
\qquad\hbox{and}\qquad
\xi\psi = {im \over \hbar} \psi,
\label{8.1}
\end{equation}
which was shown \cite{DBKP} to be strictly equivalent to the Schr\"odinger
equation on a general Newton-Cartan spacetime.
(Higher spin wave-equations can also be formulated in a similar fashion. This
has been worked out \cite{KD,Duv3} for the spin-$\frac{1}{2}$ L\'evy-Leblond
equation \cite{LL1} on a spin Bargmann manifold.)

In the $N$-body problem with metric $g_V$ and the $({\bf R},+)$-generator $\xi$ given
by (\ref{5.1},\ref{5.2}), we find that (\ref{8.1}) can be cast into the form
$$
\sum_{j=1}^N {{m \over m_j} \Delta_j \psi} + 2 \partial_t \partial_s \psi 
+ {2V \over m} \partial_s^2 \psi = 0.
$$
The second equation in (\ref{8.1}) tells that the wave function $\psi$  is
indeed of the form
\begin{equation}
\psi ({\bf q},t,s) = e^{i m s / \hbar} \Psi ({\bf q},t)
\label{8.2}
\end{equation}
where ${\bf q} = ({\bf q}_1,\ldots,{\bf q}_N)$ and $m = m_1 + \ldots + m_N$ denotes the
total mass. Hence the wave function $\Psi$ finally satisfies the $N$-body
Schr\"odinger equation
\begin{equation}
-\sum_{j=1}^N{{\hbar^2 \over 2 m_j} \Delta_j \Psi}
+ V \Psi = i \hbar {\partial \Psi \over \partial t}.
\label{8.3}
\end{equation}

Recall that on any $n$-dimensional (pseudo) Riemannian manifold $(Q,g)$ with
scalar curvature $R_g$, the operator $\Delta_g - (n-2)/(4(n-1)) R_g$ is
conformally invariant. For a Bargmann manifold with $R_g = 0$ (e.g.
when Newton field equations hold), and for $n = 3N+2$,
\begin{equation}
g^* = \Omega^2 g
\label{8.4}
\end{equation}
implies that
\begin{equation}
\Delta_g \psi = \Omega^{2+3N/2} \Delta_{g^*} \psi^*
\hskip 1 cm \hbox{whenever} \qquad
\psi^* = \Omega^{-3N/2} \psi.
\label{8.5}
\end{equation}

i) In particular, this fact entails that the extended Schr\"odinger group 
(the local diffeomorphisms $\cal C$ such that
${\cal C}^*g_0 = \Omega^2 g_0\ \&\ {\cal C}_*\xi = \xi$) ``acts'' on the
space of the solutions $\psi$ of the free Schr\"odinger equation (for $N = 1$,
say) according to
\begin{equation}
\psi\to \left({\cal C}^{-1}\right)^*\left(\Omega^{-3/2}\psi\right).
\label{8.6}
\end{equation}
Using Eqs (\ref{4.7}) and (\ref{8.2}), this ``representation'' takes the
form
\begin{eqnarray*}
\lefteqn{
\left[U\left((A,{\bf b},{\bf c},e,f,g,h)^{-1}\right)\Psi\right]({\bf q},t) =
}\hspace{.5in}\\
& & (ft + g)^{-3/2}\,\exp\left[{im\over\hbar}\left(
{f \over 2}
{(A{\bf q} + {\bf b}t + {\bf c})^2 \over ft + g} 
- \langle {\bf b},A{\bf q} \rangle  
- {t \over 2} {\bf b}^2 + h
\right)\right] \times\\
& &\Psi\left(
{A{\bf q} + {\bf b}t + {\bf c} \over ft + g},
{dt + e \over ft + g}
\right).
\end{eqnarray*}
Infinitesimally, this yields the operators \cite{Nie,Hag,BP,Hor}
\[
\left\{
\begin{array}{ll}
{\bf p} = - i\hbar\,{\displaystyle {\partial\over\partial{\bf q}}}
&\mbox{\quad (linear momentum)} \\[6pt]
{\bf L} = {\displaystyle {\bf q}\times {\bf p}}
&\mbox{\quad (angular momentum)} \\[6pt]
{\bf g} = {\displaystyle m {\bf q} - t {\bf p}}
&\mbox{\quad (boost)} \\[6pt]
H = i\hbar\,{\displaystyle {\partial\over\partial t}}
&\mbox{\quad (energy)} \\[6pt]
D = {\displaystyle  2t H - \langle {\bf q},{\bf p} \rangle + \frac{3}{2}i\hbar} 
&\mbox{\quad (dilatation)} \\[6pt]
K = {\displaystyle t^2 H - t D -{1 \over 2}m{\bf q}^2}
&\mbox{\quad (expansion)}.
\end{array}
\right.
\]

ii) In the case of a time-varying gravitational ``constant'',
$G(t) = G_0 t_0/t$ studied in Sec.~3, let $\cal D$ denote a (local)
diffeomorphism of the extended spacetime, and set $g \equiv g_{V_0}$, $g^*
\equiv {\cal D}^*g_V$. A short calculation shows that both $\psi \equiv
\psi_{V_0}$ and  $\psi^* \equiv {\cal D}^*\psi_V$ satisfy the Schr\"odinger
equation (\ref{8.1})  if $\cal D$ is a Vinti-Lynden-Bell transformation (a
straightforward generalization of (\ref{3.7}) to the $N$-body case). By
(\ref{8.2},\ref{8.5}), we find that if $\Psi_{V_0}$ is a solution of the
Schr\"odinger  equation for the $N$-body problem with $G = G_0$, then
\begin{equation}
\Psi_V({\bf q},t) =
\left({t \over t_0}\right)^{-3N/2}
\exp\Bigl(
{i\over 2\hbar t} \sum_{j=1}^N{m_j {\bf q}_j^2}
\Bigr)
\Psi_{V_0}(-{t_0 \over t}{\bf q},-{t_0^2 \over t})
\label{8.9}
\end{equation}
is a solution of the Schr\"odinger equation with the time-varying gravitational
``constant'' as given above.

Let us mention that a second-order conserved quantity associated to a
Killing tensor $\kappa$ should be quantized \cite{GRF} according to 
\begin{equation}
H_\kappa = -{\hbar^2 \over 2} \nabla_a \kappa^{ab} \nabla_b.
\label{8.10}
\end{equation}
Applied to the conformal-Killing tensor $\kappa$ given by
(\ref{6.5}--\ref{6.7}), this formula yields the following expression for the
quantized Lagrange-Laplace-Runge-Lenz vector {\bf A} in (\ref{6.8})
\begin{equation}
{\bf A} = {1 \over 2}\left({\bf L}\times{\bf p} - {\bf p}\times{\bf L}\right) 
+ m^2 m_0 G_0 {{\bf q} \over r}.
\label{8.11}
\end{equation}

\subsection{Mass prequantization}

Let us finally discuss how the mass gets quantized when the Newtonian
Taub-NUT solution (\ref{7.14},\ref{7.15}) is considered, i.e. when the fibres
of the Bargmann bundle $Q\to B$ are compact.

The basic object is the classical space of motions $(X,\omega_X)$
of our test particle of mass $m$ already introduced in Sec.~2.
As a widely accepted rule, we require that this symplectic manifold be
{\em prequantizable} \cite{JMS}. In other words, we suppose there
exists a circle bundle $Y \to X$ carrying a connection $1$-form $\vartheta_Y$
whose curvature descends to $X$ as the symplectic $2$-form $\omega_X$. It is a
well-known fact that the pre\-quantization $(Y,\vartheta_Y)$ exists iff
$\omega_X/\hbar$ defines an integral cohomology class, what in our case just
means  \begin{equation} m \ell = n {\hbar\over 2},\qquad n \in {\bf Z}.
\label{8.12} \end{equation}
Under these circumstances, the $1$-form $\vartheta_C$ of the
$8$-dimensional constrained manifold $C$ (\ref{2.4},\ref{2.5}) descends to the
discrete quotient $C_n = C/{\bf Z}_n$
obtained by taking $\psi \pmod{2\pi/n}$ where $\psi$ denotes here the
$S^1$-coordinate of the Hopf-bundle $S^3 \to S^2$ (see (\ref{7.15})).
The latter integrality  condition is, indeed, expressed as
$$
\vartheta_C\left({\partial\over\partial\psi}\right) = 2m\ell \in \hbar{\bf Z}.
$$
Thus $\vartheta_C$ is the pull-back of a $1$-form $\vartheta_{C_n}$.
Moreover, the integrable distribution
$F_n \equiv \ker(\vartheta_{C_n})\,\cap\,\ker(d\vartheta_{C_n})$
turns out to be $1$-dimensional and the quotient manifold $Y \equiv C_n/F_n$ is,
at last, our $7$-dimensional prequantum bundle carrying the connection form
$\vartheta_Y$, which is the image of $\vartheta_{C_n}$.
This construction is shown on the following diagram.
$$
\begin{array}{cccccc}
C &
\smash{\mathop{\hbox to 18mm{\rightarrowfill}}
\limits^{\scriptstyle{{\bf Z}_n}}_{\scriptstyle{}}}
& C_n
\smash{\mathop{\hbox to 18mm{\rightarrowfill}}
\limits^{\scriptstyle{\bf R}}_{\scriptstyle{F_n}}}
& Y \\
&&&
\llap{$\scriptstyle{\ker(\omega_Y)}$}\left\downarrow
\vbox to 7mm{}\right.\rlap{$\scriptstyle{S^1}$}
\\
&&& X
\end{array}
$$
Note that the prequantization condition (\ref{8.12}) provides us, in this
topologically non-trivial situation, with a {\em mass quantization} in a 
purely non-relativistic setting, i.e. purely as a consequence of non-relativistic
quantum gravity theory in which $c \to \infty$ but both $G$ an $\hbar$ are
non-zero. Mass quantization has already been discovered in the relativistic
context by Dowker and Roche \cite{DR}.


\section*{Acknowledgments}

We are indebted to G.~Burdet, M.~Perrin and F.~Ziegler for their interest and
help. Special thanks are due to H.~P.~K\"unzle for his kind permission to
include some unpublished results he had obtained in collaboration with the first
author.


\newpage

\end{document}